\documentclass[twocolumn]{aastex631}
\usepackage{natbib,amsmath}

\newcommand\gaia{\textit{Gaia}}
\newcommand\gdr[1]{\gaia~DR#1}
\newcommand\edr{\gaia~(e)DR3}
\newcommand\feh{\ensuremath{[\mathrm{Fe}/\mathrm{H}]}}
\newcommand\logp{\ensuremath{\log_{10} P}}
\newcommand\logpf{\ensuremath{\log_{10}(P_\mathrm{F})}}

\shorttitle{RR Lyrae Absolute Magnitudes}
\shortauthors{J.~Lub}

\begin{document}

\title{RR Lyrae Visual to Infrared Absolute Magnitude Calibrations\\ In the light of \gdr{3}}


\correspondingauthor{J.~Lub}
\email{lub@strw.leidenuniv.nl}

\author{K.~Looijmans}
\affiliation{Leiden Observatory, Leiden University, P.O. Box 9513, 2300 RA Leiden, the Netherlands}
\author{J.~Lub}
\affiliation{Leiden Observatory, Leiden University, P.O. Box 9513, 2300 RA Leiden, the Netherlands}
\author{A.G.A.~Brown}
\affiliation{Leiden Observatory, Leiden University, P.O. Box 9513, 2300 RA Leiden, the Netherlands}




\begin{abstract}
    A probabilistic approach has been used in combination with the parallax data from \gaia~(e)DR3 to calibrate
    Period-Luminosity-(Abundance) (PLZ) Relations covering a wide range of visual to Infrared observations of RR Lyrae
    stars.  Absolute Magnitude Relations are given, derived from the same selection of stars, for $V$, $G$, $I$,
    $K_\mathrm{s}$ and WISE $W1$ as well as for for the reddening free pseudo-magnitudes $WBV$, $WVI$ and finally also
    {\gaia} $WG$.  The classical relation between $M_V$ and \feh\ is redetermined and as an illustration distances are
    given to a few selected objects.

    \textbf{Disclaimer}: this paper reflects the presentation as given by J.~Lub at the RRLCEP2022 conference (september
    2022). Unfortunately  after preparing this report we found out that only  invited contributions would be published
    in the Proceedings.
\end{abstract}

\keywords{Stars: RR Lyrae stars, photometry, \gaia~(e)DR3, PLZ relations }


\section{Introduction}

This talk is the fourth presentation in a series given at the RR Lyrae (and Cepheid) meetings initiated in 2015 at
Visegrad (Lub, 2016, 2018, and 2021). The investigation started as an attempt to understand the origin of the $K$-$\logp$
relation and then to use this to improve absolute magnitude determinations in other photometric (visual) bands, taking
advantage of the tightness of the PL(Z) relations and the reduced interstellar absorption in the infrared.  In the
meanwhile the incredible improvements of parallax, proper motion and photometric data (\gaia\ collaboration, 2016, 2018,
2021, 2022) have made it necessary to reconsider and extend the calibrations presented at Cloudcroft in 2019 (Lub,
2021). This progress is illustrated in Table 1.~below. Much more is still to come.

It remains amusing, but not much
more should be made out of this, to note how the parallax of RR Lyrae itself (in the last column) seems to increase as
the precision of the determination increases with time.

\begin{table*}
    \caption{Time evolution of RR Lyrae parallaxes (units: milliarcseconds)}
    \label{tab:tab1}
    \begin{center}
        \begin{tabular}{lccccccc}
            \hline
            \\
            Source  & year & $V_\mathrm{average}$ & $\langle\varpi\rangle$ & $\langle\sigma_\varpi\rangle$ & $N_\mathrm{sample}$  &
            $\varpi_\mathrm{RR Lyrae} $ & References\\
            & & & mas & mas & $N_\mathrm{sample}$ & mas & \\
            \hline
            \\
            Hipparcos  & 1997 and 2007 &  11.09  &  0.992 &  3.193  &  143   &  3.46$\pm$ 0.64 & 1,2\\
            HST        & 2011      &  9.21  &  2.16  &  0.16   &  5(4)  &  3.77$\pm$ 0.13 & 3 \\
            \gdr{1} & 2016      & 11.13  &  0.938 &  0.312  &   132  &  3.64$\pm$ 0.23 & 4 \\
            \gdr{2}  & 2018      & 11.50  &  0.791 &  0.043  &   206  &    ?          & 5 \\ 
            \gaia~(e)DR3 & 2020 and 2022 & 11.50  &  0.824 &  0.021  &   207  &  3.985$\pm$ 0.027& 6,7 \\
            \gdr{4}  & (TBD)    & 11.50  &   -    &  0.010  &   207  &    -          & \\
            \gdr{5}  & (TBD)    &  &  & & &  Final catalogue & \\ 
            \\
            \hline
        \end{tabular}
    \end{center}
    {\footnotesize References: 1. Fernley et al. (1998), 2. Feast et al. (2008), 3. Benedict et al. (2011), 4,5,6 Brown
    et al. (2016), (2018), (2021), 7. Vallenari et al. (2022).}
\end{table*}

\begin{figure*}[t!]
    \includegraphics[width=\linewidth]{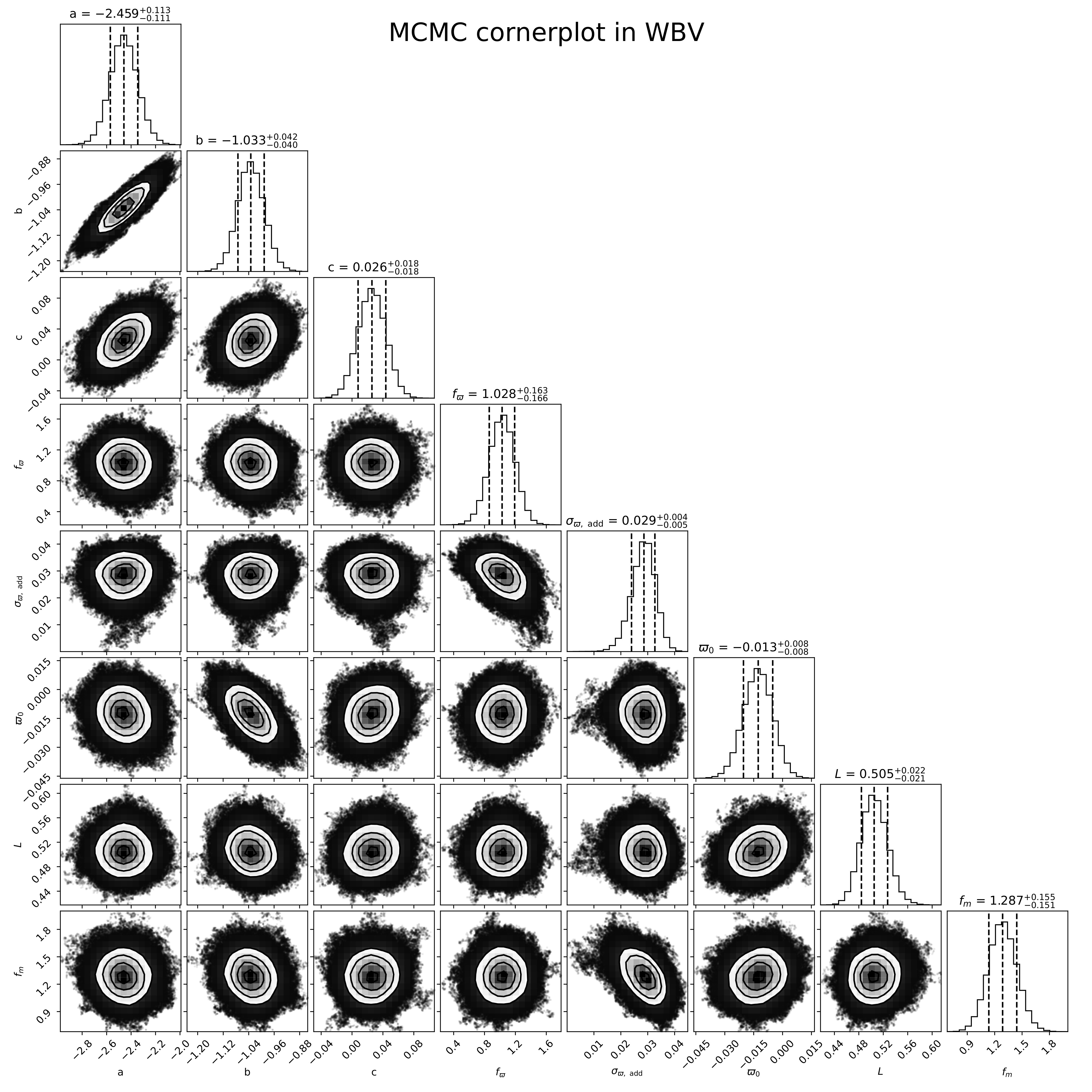}
    \caption{Cornerplot for $WBV$. Note that the coefficients $a$, $b$ and $c$ have been changed with respect to our
    definition to conform with the order in which they occur in Sesar et al.~(2017) \label{fig:corner}}
\end{figure*}

\section{ Presentation of the RR Lyrae sample}

Our sample of over 200 well studied RR Lyrae used before was updated with \edr\ parallaxes and photometry as well as W1
photometry. Monson et al.~(2017) have discussed sample of 55 brighter RR Lyrae giving us the opportunity to also add
also improved V and I photometry.  Interstellar reddening and absorption were as before based on Schlafly and Finkbeiner
(2011) with a simple correction for the pathlength within the galactic disk. An alternative approach in Muhie et al.
(2021) is based upon their $V-K$ colours. After a comparison excluding stars too close to the galactic equator and obvious
incorrect determinations (large negative absorptions from $V-K$) we could conclude that there was no offset beteeen the
two methods. 

RRc stars were fundamentalized by adding $0.1275$ to \logp\  in conformity with the value derived from the three RRd
stars in our sample and the results by Clementini et~al.~(2004) for M3. This gives as median values:
\begin{equation*}
    \logpf = -0.28\,,\\ \feh =-1.38\,, \\
    \langle V\rangle =11.50\,.
\end{equation*}

\section{ Estimating PLZ relations.}

\subsection{ First approach}

Our previous rather naive approach was to assume that in the PLZ relation:
\begin{equation*}
    M  = a + b\logpf + c(\feh + 1.35)
\end{equation*}
the coefficients $b$ and $c$ are as given by the 2015 Framework paper by Marconi et al.~(2015). Ideally photometric
parallaxes $\varpi_\mathrm{phot}$ based upon these PLZ relations can then be calculated from the fundamental relation: 
\begin{equation*}
    (m - M)_{0} = -5 \log_{10}(\varpi_\mathrm{phot}) + 10
\end{equation*}
where $m$ is the observed (pseudo)-magnitude and parallaxes are given in milliarcseconds.  The coefficient a is then
adjusted to  give a one to one relation with respect to the \gaia\ parallaxes. The \gaia\ zeropoint offset then follows
from the differences between the \gaia\ parallaxes and these so calculated photometric parallaxes. Our preliminary
calibrations based upon \gdr{2} presented in 2019 in Cloudcroft (Lub, 2021) were:
\begin{gather*}
    K_\mathrm{s} + 2.25\logpf = -1.055 + 0.18(\feh + 1.35)\notag \\
    WBV = \langle V\rangle_\mathrm{int} -3.06 \langle B-V\rangle_\mathrm{mag} \\ 
        = -1.035 -2.49\logpf\notag
\end{gather*}
Whereas the \gdr{2} zeropoint bias (see e.g.~Lindegren et al., 2018) was found as $\varpi_{0} = -0.035\pm0.010$ mas.
Please note in the appendix the change in the definition of $WBV$ used in this work.

\subsection{ A probabilistic unbiased procedure to estimate PLZ relations.}

It is clear that in this way no use is made of the full information
available, as was pointed out almost immediately when \gdr{1} (TGAS)
became available, by Sesar et al. (2017).
They introduced a probabilistic approach taking into count the full
amount of information available in the data. Following their lead one of us
(Koen Looijmans) undertook to set up and  implement a procedure to allow for
all measurement errors as well as a bias in the parallaxes $\varpi_{0}$
For the distance an exponentially decreasing volume density prior
with a scale lenghth $L$ as proposed by Bailer-Jones (2015) was adopted.

As an example we show in Fig.~\ref{fig:corner} the corner plot for the pseudo
magnitude $WBV$. 

\begin{table*}
    \caption{Summary of MCMC solutions for PLZ relations in $W1$, $K_\mathrm{s}$, $V$ ,$WBV$, $I$ and $WVI$}
    \label{tab:tab2}
    \begin{center}
        \begin{tabular}{lccccccc}
            \hline
            \\
            Band  & $a$ & $b$ & $c$ & $\varpi_{0}$ &  $L$  &    & \\
            & & & & mas & kpc &    & \\
            \hline
            \\
            $W1$  & -1.153 $\pm$ 0.042 & -2.53 $\pm$ 0.10 & 0.162 $\pm$ 0.016  & -0.0189 $\pm$ 0.0080  & 0.502 $\pm$ 0.022   &  & \\
            $K_\mathrm{s}$  & -1.088 $\pm$ 0.043 & -2.45 $\pm$ 0.11  & 0.171 $\pm$ 0.018  & -0.0149 $\pm$ 0.0080  & 0.504 $\pm$ 0.022   &   & \\
            $V$   & 0.393 $\pm$ 0.049 & -0.82 $\pm$ 0.10   & 0.273 $\pm$ 0.022  & -0.0137 $\pm$ 0.0091  & 0.506 $\pm$ 0.022  &   & \\
            $WBV$ & -1.033 $\pm$ 0.041 & -2.46 $\pm$ 0.10 &  0.026 $\pm$ 0.018  & -0.0127 $\pm$ 0.0076  & 0.505 $\pm$ 0.022  &   & \\ 

            \hline
            \\
            $I$   & -0.115 $\pm$ 0.062  & -1.22 $\pm$ 0.18 & 0.218 $\pm$ 0.025  & -0.035 $\pm$ 0.019  & 0.325 $\pm$ 0.026   &  & \\
            $WVI$   & -1.063 $\pm$ 0.055 & -2.47 $\pm$ 0.10 & 0.131 $\pm$ 0.031  & -0.027 $\pm$ 0.016  & 0.327 $\pm$ 0.026   &   & \\
            \hline
        \end{tabular}
    \end{center}
    \smallskip
    {\footnotesize The coefficients $a$, $b$ and $c$ are called $M_\mathrm{ref}$, $a$ and $b$ respectively by  Sesar et
    al. (2017), the quoted precisions are given by the average over the difference of the 84th and 16th percentile with
the median }
\end{table*}

In the summary table (Table \ref{tab:tab2}) the results for $V$ and $WVI$ are separated from the rest, because they derive
from a different source with only 55 stars.  The coefficient of the Period dependence in $W1$, $K_\mathrm{s}$, $WVI$ and
$WBV$ are all very close to $-2.50$. This might at first look surprising, but removing the effect of the interstellar
absorption also reduces at the same time the effect of temperature variations, making $WVI$ and $WBV$ very much like an
infrared magnitude, mainly measuring the stellar angular diameter.  The $I$ data also stand out because of a larger
parallax offset $\varpi_{0}$, possibly because they are brighter stars.

\section{Distance Determinations}

Armed with our Period-Luminosity relations, we (re)derive the distances to selected globular clusters and the Large
Magellanic Cloud as in Lub (2021). It should be kept in mind that pseudo-magnitudes suchs as WBV and WVI have the
unfortunate property of increasing any calibration errors in the photometry. Errors are given as mean deviations of the
mean.

First we discuss the Galactic Globular Clusters M3 and $\omega$~Cen, representative of the two  Oosterhoff groups (OI
and OII), with each over 150 RR Lyrae stars : M3 (NGC 5272) ($B$ and $V$ data Cacciari et~al.~2005, $K_\mathrm{s}$ data Bhardwaj
et~al.~2020) and $\omega$ Cen (NGC 5139) ($B$, $V$ and $K_\mathrm{s}$ data Braga et~al.~2016, 2018).  
\begin{tabular}{ll}
    \smallskip 
    M3 & $WBV$ (all stars) $(m-M)_{0} = 14.987$ ($\pm 0.059$) \\
       & $ K_\mathrm{s} $ (all stars)  $15.047$ ($\pm 0.033$) \\
    $\omega$~Cen & $WBV$ (all stars) $(m-M)_{0}=13.696$ ($\pm 0.012$)\\
                 & $K_\mathrm{s}$ (all stars)  $13.720$ ($\pm 0.004$)
     \smallskip
\end{tabular}
Unfortunately in $\omega$~Cen our preliminary result from $WVI$ falls short: $(m-M)_{0} = 13.578 $ ($\pm 0.009$). This
discrepancy remains for the moment unexplained.

In the Large Magellanic Cloud two fields, A and B, were studied in detail by Clementini et al.~(2003), di Fabrizio et
al.~(2005), and Gratton et al.~(2004), giving $BVI$ lightcurves and abundances. Proceeding as before in Cloudcroft We
derive a distance modulus $(m-M)_{0}$ of $18.539$ ($\pm 0.017$) for field A and $18.504$ ($\pm 0.017$) for field B.  The
I measurements are unfortunately very much noisier and will not be discussed any further here.  For these same two
fields, $K$ measurements were added by Muraveva et al.~(2015). More data are by Szewczyk et al.~(2008) and Borissova et
al.~(2009) for stars in the general LMC field. We derive respectively in the same order: $18.584$ for field A, $18.548$ for
field B, and $18.517$ and $18.485$ for the general field samples respectively. Errors of the median are of order
$\pm0.018$. Recently Cusano et al.~(2021) collected tens of thousands of stars with $K_\mathrm{s}$ measurements and assuming
an average $\feh = -1.50$ we find directly $(m-M)_{0} = 18.561\pm 0.052$. But this needs further investigation.

\section{The $M_V$ vs \feh\ relation}

The slope of the trend of absolute magnitude with metal abundance \feh\ in the local RR Lyrae population was once a
contentious issue, e.g.~Sandage (1993), who advocated  for a slope larger than $0.30$. However this appeared to have
been settled  to a value closer to $0.20$, based upon the  discussion of LMC data by Gratton et~ al.~(2004).
Unfortunately they did not cover a complete range of abundances. As discussed in Lub (2016, 2018, 2021), application
the $K$-\logp\ relation (and  also the $WBV$-\logp\ relation) directly leads back to the conclusion favoured by Sandage.
By calculating the absolute magnitudes with the parallaxes from the $K$-\logp\ and the $WBV$ calibrations, the afore
mentioned relation becomes (see also Muraveva et~al.~2018):
\begin{equation*}
    M_V = 0.624(\pm 0.008) +  0.334(\pm 0.015) (\feh + 1.35)
\end{equation*}
A mean value of $M_V = 0.60$ has of course been in use for a long time.

\section{ \gaia\ magnitudes: $WG$}

The photometry from the \gaia\ Satellite was originally not considered for this research. The reason for this was the
fact that the published data are derived as straight means over the measured intensities, without reference to their
phase in the lightvariation. However it clear as is true for WBV the combination from the \gaia\ magnitudes $G$,
$G_\mathrm{BP}$ and $G_\mathrm{RP}$, viz. 
\begin{equation*}
    WG = G - 1.85 (G_\mathrm{BP}-G_\mathrm{RP})
\end{equation*}
is a reddening free pseudo-magnitude derived from simultaneously measured intensities, which will reduce the lightcurve
amplitude in a similar way as for $WBV$.

This is indeed borne out by comparing $WG$ with $W1$,$K_\mathrm{s}$ and $WBV$. Somehow taking out most of the
temperature dependence gives rise to a kind of pseudo IR magnitude mainly dependent on the angular diameter as mentioned
before. Apart from having a slope indistinguishable from one the median scatter is  $0.043$, $0.056$ and $0.072$ (rms
differences $0.056$, $0.073$ and $0.102$) respectively. This is indicative of the precision of the photometric
measurements, where apparently W1 is superior.

As a shortcut we have used our calibrations for $WBV$ and $K_\mathrm{s}$ to predict each stars' photometric parallax and
taking the average value. A simple least squares approach, which of course then no longer takes into account the
actually measured uncertanties and priors in the full probablistic approach the gives us:
\begin{equation*}
    WG = -1.093 - 2.475 \logpf + 0.121 (\feh + 1.35)
\end{equation*}
The errors on the coefficients are respectively: $0.020$, $0.065$ and $0.010$ and with this choice of $WG$ the \gdr{3}
zeropoint bias $\varpi_{0}$ comes out as $-0.008$  with an rms of $0.045$ ($0.0032$ for the median).  A comparison with
the separate investigation of Garofalo et al.~(2022) shows very good agreement for coeficients and zeropoint.

It will be of interest to see how the full Bayesian approach will change this result when it is finally done. At any
rate this calibration is fully consistent with our results for the other bands.

\section{Conclusions}

\gaia\ has set the zeropoint of the RR Lyrae Period-Luminosity relations with great precision. Here we have
presented a set of Period-Luminosity-(Abundance) relations, which are internally consistent over the range fom $V$ to
$W1$, because they are based on the same stellar sample. With only a few variables very good distances can de determined
to all objects in the Local Group, which contain RR Lyrae stars. 

\begin{acknowledgements}
RRLCEP2022 at La Palma was a great opportunity for a re encounter after more than two difficult years. We wish to
thank the organizers, who unfortunately could not know that the conference would be delayed by an approaching
tropical storm causing epic delays in arrivals on the island.  This work was done by Koen Looijmans as a Bachelor
research project as the final requirement for his BSc examination.  J. Lub wishes to thank the Leiden Kerkhoven
Bosscha Fund and Sterrewacht Leiden for travel support to La Palma.  Extensive use was made of the data from the
European Space Agency (ESA) mission \gaia\  (\url{ http://cosmos.ea.int/gaia}) processed by the
\gaia\ Data Processing and Analysis Consortium DPAC (\url{http://cosmos.esa.int/web/gaia/dpac/consortium}).  Funding
for the DPAC has been provided by the national institutions, in particular the institutions participating in the
\gaia\ Multilateral Agreement. This publication makes use of data products from the Wide-field Infrared Survey
Explorer, which is a joint project of the University of California, Los Angeles, and the Jet Propulsion
Laboratory/California Institute of Technology, funded by the National Aeronautics and Space Administration
\end{acknowledgements}

\appendix
\section{Intensity versus magnitude averages.}

RR Lyrae stars have large $V$ amplitudes up till $1.25$ mag for RRab stars. The ratios between Blue ($B$) and near
Infrared ($I$) amplitudes $Ab$ and $Ai$ to $V$ amplitude $Av$ are $1.29$ and $0.63$ respectively.  Published data are a
mixture of mean intensity converted to magnitude and average magnitude. Sometimes even intensity ratios are averaged.
From our large collection of fully covered lightcurves (Lub, 1977) we derived :

\begin{equation*}
    \langle V\rangle_\mathrm{int} - \langle V\rangle_\mathrm{mag} = -0.047 Av^{2}
\end{equation*}

The $B$ and near infrared $I$ lightcurves are very similar to the $V$ lightcurves so
we can just substitute $Ab=1.29Av$ or $Ai=0.63Av$ for $Av$. 
\begin{gather*}
    (\langle B\rangle_\mathrm{int}-\langle V\rangle_\mathrm{int}) - (\langle B\rangle_\mathrm{mag}-\langle
    V\rangle_\mathrm{mag}) =-0.031Av^{2}\\
    (\langle V\rangle_\mathrm{int} - \langle I\rangle_\mathrm{int}) -(\langle B\rangle_\mathrm{mag} - \langle
    V\rangle_\mathrm{mag}) = -0.027Av^{2}
\end{gather*}

In this way also different definitions of the pseudo magnitude WBV can be
related. However since our earlier work we have changed the definition of
$WBV$ to conform with Neeley et al. (2019) replacing the absorption correction
with 3.128 instead of 3.06, viz.
\begin{equation*}
    WBV = \langle V\rangle_\mathrm{int} - 3.128\langle B-V\rangle_\mathrm{mag}
\end{equation*}
because we consider $\langle V\rangle_\mathrm{int}$ a good estimate of the bolometric luminosity
and $\langle B-V\rangle_\mathrm{mag}$ a better temperature indicator than the intensity average.
However over our sample the mean of $ Av^{2} =0.980 $  so using the definition
of $WBV$ preferred by Marconi et al. would cause a positive shift of 0.094 in
the zeropoint of the WBV relation.
Lightcurve amplitudes are increasing towards shorter periods. After some
experiments we have concluded that this increases the slope
when only RRab stars are considered in deriving the PLZ relation.
The coefficient of the period dependence increasing from close to
$-2.5$ to $-2.75$.

For $WVI$ and also for \gaia\ $G$, $G_\mathrm{BP}$ and $G_\mathrm{RP}$ we have prefered to use the published
intensity means:
\begin{gather*}
    WVI = \langle I\rangle_\mathrm{int} -   1.38(\langle V\rangle_\mathrm{int}  -  \langle I\rangle_\mathrm{int})\\
    WG = \langle G\rangle_\mathrm{int} - 1.85(\langle G_\mathrm{BP}\rangle_\mathrm{int} - \langle G_\mathrm{RP}\rangle_\mathrm{int})
\end{gather*}

\end{document}